\documentclass{aipproc}

\usepackage{graphicx}
\usepackage{natbib}

\layoutstyle{6x9}
\classification{}

\begin{document}

\title[Understanding Black Hole Formation]{Black Hole Formation in X-Ray Binaries: \\ The
Case of XTE J1118+480} 

\author{T. Fragos}{address={Northwestern University, Department of Physics and Astronomy, 2145 Sheridan Road, Evanston, IL 60208, USA}, email={tassosfragos@northwestern.edu}}

\author{B. Willems}{address={Northwestern University, Department of Physics and Astronomy, 2145 Sheridan Road, Evanston, IL 60208, USA}, email={b-willems@northwestern.edu}}

\author{N. Ivanova}{address={Canadian Institute for Theoretical Astrophysics, University of Toronto, 60 St George, Toronto, ON M5S 3H8, Canada}, email={nata@cita.utoronto.ca}}

\author{V. Kalogera}{address={Northwestern University, Department of Physics and Astronomy, 2145 Sheridan Road, Evanston, IL 60208, USA}, email={vicky@northwestern.edu}}


\begin{abstract}

In recent years, an increasing number of proper motions have been
measured for Galactic X-ray binaries. When supplemented with accurate
determinations of the component masses, orbital period, and donor
luminosity and effective temperature, these kinematical constraints
harbor a wealth of information on the systems' past evolution. The
constraints on compact object progenitors and kicks derived from this
are of immense value for understanding compact object formation and
exposing common threads and fundamental differences between black hole
and neutron star formation. Here, we present the results of such an
analysis for the black hole X-ray binary XTE J1118+480. We present results from modeling the mass
transfer phase, following the motion in the Galaxy back to the birth
site of the black hole, and examining the dynamics of symmetric and
asymmetric core-collapses of the black hole progenitor.


\end{abstract}

\keywords{Stars: Binaries: Close, X-rays: Binaries,
  X-rays: Individual (XTE\,J1118+480)} 

\maketitle

\section{INTRODUCTION}

In recent years the observed sample of Galactic black-hole (BH) X-ray binaries (XRBs) has increased significantly. For many of these systems there exists a wealth of observational information about their
current physical state: BH and donor masses, orbital period, donor's position on the H-R
diagram and surface chemical composition, transient or persistent and Roche-lobe overflow
(RLO) or wind-driven character of the mass-transfer (MT) process, and distances.
Lately the full proper motion has been measured for a handful of these systems
(\citet{Mirabel2001}, \citet{Mirabel2002}, \citet{Mirabel2003}), which in addition to the earlier
measurements of center-of-mass radial velocities gives us information about the
3-dimensional kinematic properties of these binaries. The vast amount of observational work that has been done provides us with a unique opportunity to study and try to understand the formation and evolution of black holes in binaries. 

\citet{Willems2005} showed how using all the currently available observational constraints, one can  uncover the evolutionary history of the system from the present state back to the time just prior to the core-collapse event, and they applied their analysis to the BH XRB GRO\,J1655-40. In the work presented here we follow the same framework but we focus on the case of XTE J1118+480. 

%

\section{OBSERVATIONAL CONSTRAINTS FOR XTE\,J1118+480}

XTE J1118+480 is one of the X-ray novae that has been dynamically confirmed to contain a black hole primary and the first to be located at a high galactic latitude (\citet{Remillard2000}, \citet{Cook2000}, \citet{Uemura2000}). After the first detection several groups performed detailed observation in order to unravel the properties of the binary system (\citet{McClintock2001}, \citet{Wagner2001}, \citet{Gelino2006}). The proper motion of XTE J1118+480 was observed with the VLBA by \citet{Mirabel2001}. More recent observations with the 10-m KECK II telescope (\citet{Gonzalez2006}) revealed that the secondary star has a supersolar surface metallicity $[Fe/H]=0.2 \pm 0.2$. The observational constraints that we will use in our analysis are summarized in Table 1.

\begin{table}[ht]
\begin{tabular}{lccc}
\hline

 \tablehead{1}{l}{b}{Parameter} & 
 \tablehead{1}{c}{b}{Notation} & 
 \tablehead{1}{c}{b}{Value}&
 \tablehead{1}{c}{b}{Reference
\footnote{(1) \citet{Remillard2000}, (2) \citet{McClintock2001}, (3) \citet{Wagner2001}, (4) \citet{Mirabel2001}, (5) \citet{Gelino2006}, (6) \citet{Gonzalez2006}}
} \\
\hline

Distance                    & $d$           & $1.85 \pm 0.36$\,kpc & (2), (3), (4) \\
Galactic longitude          & $l$           & $157.7^\circ$ &  (1) \\
Galactic latitude           & $b$           & $+62.3^\circ$   &  (1) \\
Velocity towards the Galactic Center\footnote[2]{relative to the local standar of rest}       & $U$ & $-105 \pm 16$\,km\,s$^{-1}$ & (4) \\
Velocity in the direction of the Galactic Rotation$^\dag$ & $V$ & $-98 \pm 16$\,km\,s$^{-1}$   & (4) \\
Velocity towards the Northern Galactic Pole$^\dag$& $W$ & $-21 \pm 10$\,km\,s$^{-1}$     & (4) \\
Orbital Period              & $P_{\rm orb}$  & $0.17 \pm 0.001$\,days & (2), (3) \\
Black Hole Mass             & $M_{\rm BH}$   & $8.0 \pm 2.0\,\rm{M_\odot}$  & (2), (3), (5) \\
Donor Mass                  & $M_2$          & $0.455 \pm 0.355\,\rm{M_\odot}$  & (2), (3), (5) \\
Donor Luminosity            & $L_2$          & $0.0594 \pm 0.0319\,\rm{L_\odot}$ & (2), (3) \\
Donor Effective Temperature & $T_{\rm eff2}$ & $4409 \pm 440$\,K       &  (2), (3) \\
Donor Surface Metallicity   & $[Fe/H]$       & $0.2 \pm 0.2$           &  (6) \\
\hline
\end{tabular}

\caption{Properties of XTE\,J1118-480. }
\label{1118param}

\end{table}

\section{Outline of analysis methodology}

The formation of a low mass X-ray binary from a primordial binary in the galactic field requires, according to our current understanding,  an extreme initial mass ratio and a mass for the primary star of at least $20-25\,\rm{M_\odot}$ so that the formation of a BH is possible. If the initial orbit is tight enough, with period less than $10\, \rm{yr}$, the primary star soon overflows its Roche lobe (RLO ) and, during an unstable common envelope phase, looses most of its hydrogen-rich envelope. If the energy dissipated is enough to expel the whole envelope and the binary avoids a merger, then at the end of the common envelope phase we are left with the helium core of the primary star and an unevolved main sequence secondary star in a tight orbit. Subsequently the helium core collapses into a BH imparting possibly an asymetric natal kick on the system. Eventually stellar evolution and/or orbital angular momentum losses cause the secondary in its turn to fill its Roche lobe and transfer mass to the BH, forming a BH XRB.

In this paper we restrict ourselves to the formation of XTE J1118+480 through the above mentioned "standard" channel assuming that the system was born in the galactic disk and thus the donor has solar metallicity. For a discusion of a possible globular cluster origin of the system we refer to Fragos et al. 2006 (in preparation). 

The analysis we follow in order to trace back the evolutionary history of XTE J1118+480 incorporates a number of calculations which can be summarized in four discrete steps.

\begin{figure}[h]
\includegraphics[width=0.5\textwidth]{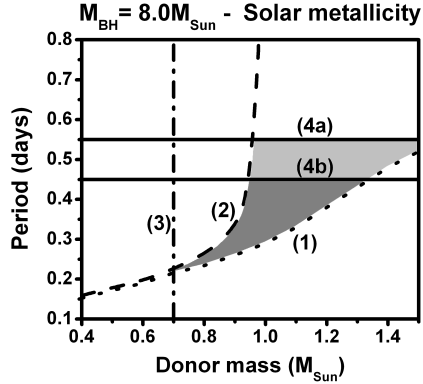}
\caption{Constraints on the parameter space. See text for explanation.}
\end{figure}

We first use a stellar and binary evolution code (\citet{Podsiadlowski2002}, \citet{Ivanova2003}, \citet{Kalogera2004}) and calculate a grid of evolutionary sequences for binaries in which a BH is accreting mass
from a Roche-lobe filling companion. The parameter space (initial donor mass, BH mass, and period at the start of RLO) that we have to cover with this grid is finite and limited by a number of constraints (see figure 1). The initial Roche lobe radius at the onset of the mass transfer (MT) has to be larger than the ZAMS radius of the donor. This sets a minimum initial period for given masses of the two components (dotted line 1). We can also set a maximum initial period by requiring that the donor fills its RL within a Hubble time (dashed line 2). A minimum mass for the donor star is set by the fact that the secondary star, at its RLO, should be hotter and more luminous than the currently observed donor, so that it can eventually reach the observed temperature and luminosity (dashed-dotted line 3). Finally, given that the observed period is below the bifurcation period, the period at RLO should also be below the bifurcation period. The value of the bifurcation period depends on the masses of the system and the prescription of the magnetic braking that we apply (solid line 4a for \citet{RVJ1983} prescription and solid line 4b for \citet{IvanovaTaam2003} prescription). 

 To consider the full range of possibilities, we include sequences for both conservative (for
sub-Eddington rates) and fully non-conservative MT. For each sequence,
we examine whether at any point in time the calculated binary
properties are in agreement with the
observational measurements. From the successful sequences we derive the properties of the
binary at the onset of the RLO phase: initial BH and donor masses,
orbital period, and age of the donor star. The time at which the fully
successful sequences satisfy all observational constraints furthermore
provides an estimate for the donor's current age.

The second step is to follow the orbital evolution of the system due to tides
and gravitational radiation prior to the XRB phase. This calculation yields the the post-collapse semi-major axis and eccentricity of the system.

Next, we consider the kinematic evolutionary history of the XRB in the
Galactic potential. In particular, we use the current position and the
measured 3D velocity with their associated uncertainties to trace the
Galactic motion back in time. Combined with the tight constraints
on the current age of the system given by the successful MT sequences
this allows us to determine the position and velocity of the binary at
the time of BH formation (we denote these as the ``birth'' location
and velocity). By subtracting the local Galactic rotational velocity
at this position from the system's total center-of-mass velocity, we obtain an estimate for the {\em peculiar} velocity of the binary
right after the formation of the BH.

Finally, knowing all the properties of the system just after the super nova (SN), we analyze the dynamics of the core collapse event due to mass loss and possible natal kicks. Based on angular momentum and energy conservation we derive constraints on the pre-SN binary properties (BH progenitor mass and orbital separation) and the natal kick (magnitude and direction) that may have been imparted to the BH.

\section{Progenitor constraints}

Putting together the different pieces of our analysis we can uncover the complete picture of the evolutionary history of XTE J1118+480 back to the time of BH formation. The study of the MT sequences strongly constrains the properties of the binary at the start of its XRB phase. We find
that at the start of RLO the donor mass is 1.2-1.3\,$M_\odot$, the BH mass
is 6-10\,$M_\odot$, and the orbital period 0.5-0.6\,days.

\begin{figure}[h]
   \includegraphics[width=0.585\textwidth]{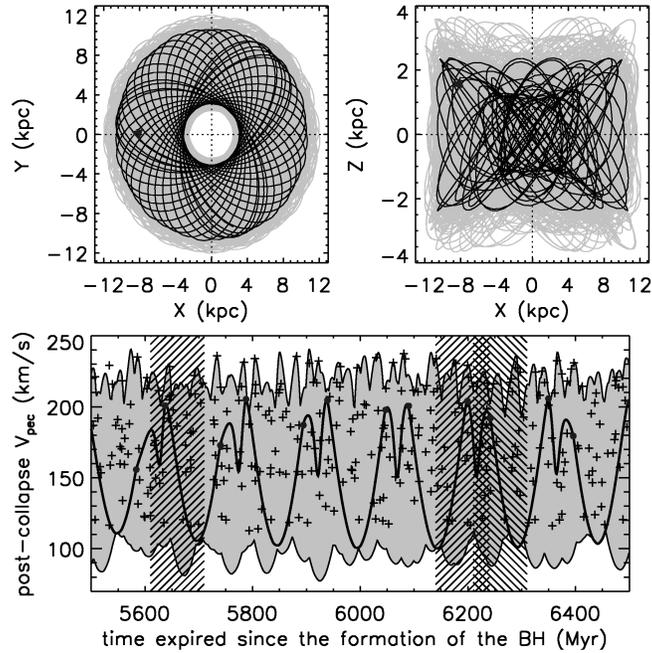}
   \caption{Upper panel:Orbit of XTE J1118+480 in the Galaxy in the XY- and XZ-planes. The grey lines indicate the uncertainties in the orbit associated with the
    error bars in the distance and the velocity components.
    Lower panel: Post-SN peculiar velocity as a function of
  the time expired since the formation of the BH. The light grey area
  indicates the uncertainties resulting from the error bars in the
  distance and the velocity components. Possible birth times of the BH given by the successful
  MT sequences are indicated by the hatched vertical bars. The crosses indicate crossinges of the orbit with the galactic plan.}
\end{figure}

\begin{figure}[h]
\includegraphics[width=0.45\textwidth]{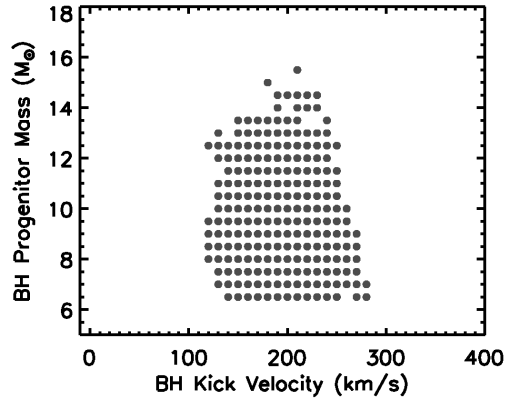}
\caption{Constraints on the mass $M_{\rm He}$ of the BH's helium star
  progenitor and the magnitude $V_kick$ of the kick velocity imparted to
  the BH.}
\end{figure}

Our kinematic study shows that the system follows a semiperiodic motion crossing the galactic plane about every 85\,Myr. The motion is bound in the galactic potential covering values of R between 3 and 12 Kpc while oscillating in the Z direction with a maximum amplitude of 3.5\,kpc. The post-SN peculiar velocity of the system spans a wide range of values from 80 to 240\,$\rm{km\, s^{-1}}$ (see Figure 2). In order to identify the post-SN peculiar velocity of the system, we search for the crossing of the galactic plane that occurred during the time interval of 50\,Myr before and after the current age of the system. Taking also into account the error bars in the current velocity and position of the system we can derive an upper and lower limit for the post-SN peculiar velocity of the system. For all the successful MT sequences, we find $120\,\rm{km\, s^{-1}}<V_{\rm{pec,postSN}}<240\,\rm{km\, s^{-1}}$.

To understand the core-collapse event leading to the formation of the BH, we are
mainly interested in the constraints derived for the mass of the BH's helium star progenitor
and the kick velocity that may have been imparted to the BH at birth. Using the set of possible solutions that we derived for the system parameters exactly after the BH formation, we constrain the BH-progenitor mass and the possible BH kick magnitude. We find that the mass of the helium star is constrained to $6.5\,\rm{M_\odot} \leq M_{\rm{He}} \leq 15.5\,\rm{M_\odot}$ and that an asymetric natal kick is required for the formation of the system. The magnitude of the kick should be in the range $120\,\rm{km\, s^{-1}} < V_{\rm{kick}}<280\,\rm{km\, s^{-1}}$ (see Figure 3).

\section{Discussion}
In this paper we constrained the progenitor properties and the formation of the BH in the XRB XTE\,J1118+480 assuming that the system originated in the Galactic disk and the donor had solar
metallicity. In line with previous investigations, we find that a high magnitude asymetric natal kick is not only plausible but required for the formation of the system. The
minimum kick is significantly larger than that of the BH in GRO\,J1655-40
where the natal kick could be as low as a few tens of km/s.

A globular cluster origin of XTE\,J1118+480 requires a low metallicity donor star. Study of such MT sequences show that the ones that successfully reproduce the observational properties usually have MT rates above or marginally below the critical MT rate that separates transient from persistent systems. Furthermore the current super solar surface metallicity of the donor star requires that the star was in a binary system earlier in its life where the SN explosion of the primary star enhanced its surface with metals. Such a scenario involves a lot of fine tuning and seems rather improbable.   

In subsequent analyses, we intend to apply the procedure outlined
above to a number of XRBs:
Cyg X-1, LS\,5039, LSI\,$+61^\circ\,303$, Vela\,X-1, 4U1700-37, Sco\,X-1. 
By examining both NS and BH systems and both
RLO and stellar wind induced MT, we hope to unravel the systematic
dependencies between the masses of newly formed compact objects and
their immediate pre-SN progenitors, the mass lost at core collapse,
and the possible kick velocity magnitude imparted to the compact
object at birth. 

\def\acknowledgments{%
\section*{ACKNOWLEDGMENTS}
}

\acknowledgments 

We are indebted to Laura Blecha for sharing the code used to follow
the motion of XTE\,J1118+480 in the Galactic potential. This work is
supported by a Packard Fellowship in
Science and Engineering grant and an NSF CAREER award to VK

\bibliography{fragos_1}

\end{document}